\documentclass[prb,twocolumn,showpacs,preprintnumbers,amsmath,amssymb]{revtex4}
\usepackage{graphicx}
\usepackage{amsmath}
\usepackage{amsbsy}
\usepackage{dcolumn}
\usepackage{bm}
\usepackage{color}
\def\rr{{\bf r}}
\def\rr{{\bf r}}

\def\k{{\bf k}}

\def\q{{\bf q}}

\def\rr{{\bf r}}

\begin{document}

\title{Sublattice model of atomic scale pairing inhomogeneity in a
superconductor}

\author{Vivek Mishra}
\email{vivekm@phys.ufl.edu}
\affiliation{Department of Physics, University of Florida,
Gainesville, Florida 32611-8440, USA}

\author{P. J. Hirschfeld}
\email{pjh@phys.ufl.edu}
\affiliation{Department of Physics, University of Florida,
Gainesville, Florida 32611-8440, USA}

\author{Yu. S. Barash}
\affiliation{Institute of Solid
State Physics, Russian Academy of Sciences, Chernogolovka, 142432
Russia}

\date{\today}

\begin{abstract}
{We study a toy model for a superconductor on a bipartite lattice,
where intrinsic pairing inhomogeneity is produced by two different
coupling constants on the sublattices. The simplicity of the model
allows for analytic solutions and tests of the consequences of
atomic-scale variations in pairing interactions which have been
considered recently in the cuprates.  We present results for the
transition temperature,  density of states, and thermodynamics of
the system over a  phase diagram in the plane of two pairing
coupling constants.  For coupling constants of alternating sign, a
gapless superconducting state is stable.   Inhomogeneity is
generally found to enhance the critical temperature, and at the
same time the superfluid density is remarkably robust: at $T=0$,
it is suppressed only in the gapless phase.}
\end{abstract}

\pacs{74.81.-g,74.62.-c,74.78.Na,74.25.Dw}

\maketitle

\section{Introduction }
In recent years, the effect of underlying inhomogeneities in
superconductors has attracted many researchers. In the context of
high temperature superconductors (HTSC),  checkerboard local
density of states (LDOS) oscillations and strong nanoscale gap
inhomogeneity have been observed in scanning tunnelling
spectroscopy (STS) experiments \cite{Fischerreview}, and signals
in dynamical susceptibility measured by neutron scattering have
been interpreted as stripelike nanoscale modulations of charge and
spin degrees of freedom \cite{Tranquadareview}. While it is still
not clear whether these modulations are intrinsic in nature or
driven entirely by disorder, others have pointed out that the very
existence of inhomogeneity may  enhance $T_c$, and implicitly
suggested that the phenomenon of high critical temperatures may
rely on it \cite{Kivelson,Dagotto}.  In particular, Martin et al.
\cite{Kivelson} studied a simple model, in which they included a
1D inhomogeneous pair potential varying on an arbitrary length
scale $\ell \gtrsim a$, where $a$ is the lattice spacing, and
argued for an increase of the critical temperature with
inhomogeneity, with a maximum enhancement obtained for $\ell\simeq
\xi_0$, where $\xi_0$ is the coherence length of the analog
homogeneous system. Similar results were obtained by Loh and
Carlson, where they found however that an increase in the
transition temperature occurred at the cost of reduced superfluid
density \cite {loh}. Aryanpour et al. \cite{scalettar} performed a
systematic numerical study of $s$-wave superconductivity on
checkerboard, stripe and random patterns of pairing potentials in
an attractive Hubbard model; they also found a positive
correlation between critical temperature and inhomogeneity for
many situations.  It is also known that models with enhanced $T_c$
and significant inhomogeneity are associated with unusual features
in electronic structure.
For example, recent calculations of high-quality ultrathin s-wave
superconducting films have shown that the surface-driven
inhomogeneity can enhance $T_c$ due to weakly dispersive
low-energy subgap states in the system\cite{Barash}.  The above
studies were primarily considerations of BCS type models.
Recently, however, Tsai et al.\cite{Tsai_etal} have studied small
size repulsive Hubbard models  with a ``checkerboard"-like
modulation of hopping matrix elements using exact diagonalization,
and also concluded that the inhomogeneous models could lead to
pairing at very high binding energy scales.

 Nunner et al. \cite{Nunner} underlined the importance of
this phenomenon for cuprates by explaining many of the
correlations between the measured local gap and other observables
in Bi-2212 by assuming a $d$-wave pairing interaction $g(\rr)$
varying on a unit cell size, and driven by dopant disorder. If the
effective pair interaction  can vary on an atomic scale, and the
BCS coherence length is sufficiently short, one can imagine
situations where in a complex periodic crystal with large unit
cell the pairing interaction varies over an atomic length scale in
the absence of disorder, i.e. a periodic, modulated $g(\rr)$,
which will give rise to a periodic, modulated order parameter
$\Delta(\rr)$. The properties of such a system are important to
understand, because it represents the simplest type of nanoscale
pairing modulation, and one may study the relationship between
superconducting properties and inhomogeneity easily.  Aryanpour et
al. \cite{scalettar} considered systems of this type within a
purely numerical approach, and obtained a variety of interesting
results. Nevertheless, in such a treatment it is sometimes
difficult to extract simple conclusions and understand exactly
which aspects of superconductivity are really enhanced or
suppressed by inhomogeneity, and why.

In this paper, we consider the simplest such model, a bipartite
lattice in two dimensions with two different values of $g$ on two
interpenetrating sublattices.   The homogeneous version of the
model has a superconducting ground state on a bipartite lattice in
arbitrary dimension and filling \cite{campbell}.  Here we
calculate the ground state and thermodynamics of this system for
any values $(g_A,g_B)$, where $A$ and $B$ represent the two
sublattices.  We distinguish however two cases: one, where both
$g$'s have the same sign; and a second, where they differ.
Situations like the latter case where the pairing potential
changes sign abruptly over a unit cell distance have been
considered in the impurity context
\cite{Chattopadhyay,Andersen_phaseimpurity}, but not in periodic
cases, and it is not clear whether true superconducting solutions
can exist in such a situation.

\begin{figure}
\includegraphics[width=0.45\textwidth]{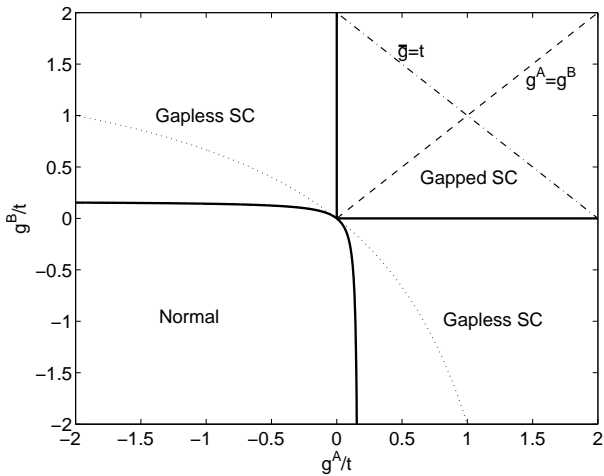}
\caption{$T=0$  mean field phase diagram  in the space of two
coupling constants $g_A$ and $g_B$ found in this paper.   The
solid lines
 represent  transitions between normal metal (defined by $\Delta_A= \Delta_B=0$),
   gapped and  gapless superconducting phases.
 The  dashed-dotted
line represents the line of constant average pairing $\bar g=t$
studied in the text.  The dashed line represents the homogeneous
BCS case with $g^A=g^B$.  Finally, the dotted line represents the
normal-gapless superconducting transition line as calculated
analytically within the approximation described in the text using
the ``window" density of states (wDOS). } \label{fig:phsd}
\end{figure}

In section II, we introduce the sublattice pairing model and give
exact expressions for the order parameter and critical temperature
$T_c$ within the generalized BCS scheme.  We show that $T_c$ can
indeed be enhanced by detuning the coupling constants from the BCS
point $g_A=g_B$. We then derive in section III the quasiparticle
excitations and thermodynamic properties of the model.  When the
average pairing strength is held constant but inhomogeneity is
increased, the model displays a shrinking  spectral gap in the total
DOS. In the case of opposite sign coupling constants, the excitation
spectrum becomes gapless, and the mean field solution corresponds to
an order parameter which oscillates in sign from point to point. In
section IV, we calculate the superfluid density $n_s$ for the model,
and show that for the case of same-sign coupling constants
inhomogeneity preserves a robust $n_s$ at $T=0$.  In the case of
opposite signs, however, $n_s$ is suppressed by the onset of a
residual normal fluid. The variations of superconducting
phenomenology found in our solution to the model is summarized in the
phase diagram shown in Fig. \ref{fig:phsd}.

\section{ Model} 
\begin{figure}
  \begin{center}
  \includegraphics[width= 4.5 cm,height=4.5 cm]{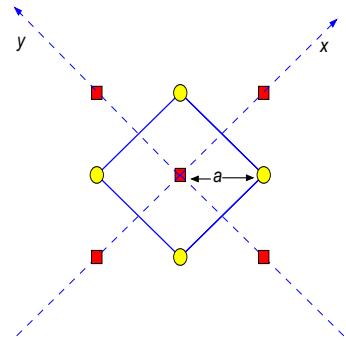}
   \caption{Unit cell of  square lattice, where squares and circles denote sites of type
   "$A$" and "$B$",  and  the lattice  spacing is denoted by $a$, which we have set to 1 throughout our discussion.
   The coordinate system used in Fourier transformation is shown with two dashed lines. }
  \label{fig:BZ}
  \end{center}
\end{figure}

The Hamiltonian is given by Eq. (\ref{eq:mdl}),
where $c_i,c^{\dagger}_i$ are the electron annihilation and
creation operators. $t$ is the strength of hopping energy between
the nearest neighbors and $\Delta_i$ is the mean field gap at site
$i$. The sum in the Hamiltonian is over all sites and spins
denoted by $\sigma$. The unit cell of the bipartite lattice is
shown in Fig. \ref {fig:BZ}, and the Fourier transformation in Eq.
(\ref{eq:ft}) is defined with respect to this cell. We first
consider the case of half-filling, for which the chemical
potential $\mu$ is set to zero.

\begin{equation}
H=\sum_{i,\delta,\sigma}-tc^{\dagger}_{i,\sigma}c_{i+\delta,\sigma}+
\Delta_{i}c^{\dagger}_{i,\sigma}c^{\dagger}_{i,-{\sigma}}.
\label{eq:mdl}
\end{equation}
In anticipation of the doubling of the unit cell due to an
interaction specified on each sublattice, we Fourier transform
fermionic operators as
\begin{equation}
c_{i,\sigma}=\sum_\k c_{\k,\sigma}e^{i\k\cdot \rr_{i}},
\label{eq:ft}
\end{equation}
where the sum now runs over the arising for staggered ordering
(``antiferromagnetic'')
half-Brillouin zone specified by $0\le |k_i| \le \pi/(\sqrt{2}a)$
for $i=x,y$, with directions specified in Fig. \ref{fig:BZ}.  The
Hamiltonian now becomes
\begin{eqnarray}
H&=&\sum_{\k,\sigma}\xi_k c_{\k\sigma}^{\dagger
A}c^{B}_{\k\sigma}+ \Delta^{A}c^{\dagger A}_{\k\sigma}c^{\dagger
A}_{\k-{\sigma}}+ \Delta^{B}c^{\dagger B}_{\k\sigma}c^{\dagger
B}_{\k-{\sigma}}
\\ &=& \sum_\k \tilde c_\k^\dagger \cdot M \cdot \tilde c_\k,
\label{eq:hk}
\end{eqnarray}
where we have assumed $s$-wave pairing and the dispersion relation
is given by
\begin{equation}
\xi_{k}=-4t \cos(\frac{k_{x}}{\sqrt{2}})
\cos(\frac{k_{y}}{\sqrt{2}}). \label{eq:disp}
\end{equation}
The matrix $M$ may be expressed as
\begin{equation}
M=\left[\begin{array}{cccc}
0  & \xi_{k} & \Delta^{A} & 0 \\
\xi_{k} &  0 & 0  & \Delta^{B} \\
\Delta^{A} & 0 & 0 & -\xi_{k}\\
0 & \Delta^{B} &  -\xi_{k}& 0\\
\end{array}\right]
\label{eq:mat}
\end{equation}  in the staggered particle-hole space spanned by
$\tilde c_\k
=(c^A_{-\k\sigma},c^B_{-\k\sigma},c^{A\dagger}_{\k-{\sigma}},c^{B\dagger}_{\k-{\sigma}})$.
After performing the canonical transformation
\begin{equation}
c^{\dagger
\alpha}_{\k\sigma}=u^{*\alpha}_{n}(\k)\gamma^{\dagger}_{\k\sigma}+\sigma
v^{\alpha}_{n}(\k)\gamma_{\k,{-\sigma}}, \label{eq:ctran}
\end{equation} with $\alpha=A,B$,
we find the quasiparticle energies $\pm E_{1,2}$, with
\begin{equation}
E_{1,2}=\frac{\mp\Delta^{A}\pm\Delta^{B}+
\sqrt{\left(\Delta^{A}+\Delta^{B}\right)^{2}+4\xi_\k^{2}}} {2},
\label{eq:E1}
\end{equation} which reduce to the usual $E_\k=\sqrt{\xi_\k^2+\Delta^2}$
for $\Delta^\alpha\rightarrow\Delta$.
%
%

The eigenstates of the problem, of the form
$|u_A,u_B,v_A,v_B\rangle$, are given by
\begin{align}
&|\lambda_{1},-E_{1}>=[-x_{1},-x_2,-x_{1},x_2]
{\frac{1}{\sqrt{2\left(x^2_1+x^2_2\right)}}} \label{eq:v1},\\
&|\lambda_{2},-E_{2}>=[-x_{2},-x_1,x_{2},-x_1]
{\frac{1}{\sqrt{2\left(x^2_1+x^2_2\right)}}} \label{eq:v2}, \\
&|\lambda_{3},E_{2}>=[-x_{2},x_1,-x_{2},-x_1]
{\frac{1}{\sqrt{2\left(x^2_1+x^2_2\right)}}} \label{eq:v3}, \\
&|\lambda_{4},E_{1}>=[-x_{1},x_2,x_{1},x_2]
{\frac{1}{\sqrt{2\left(x^2_1+x^2_2\right)}}} \label{eq:v4},
\end{align}
where
\begin{equation}
x_{1,2}=\sqrt{\sqrt{\left(\Delta^{A}+\Delta^{B}\right)^{2}+4\xi^{2}}
\mp\left(\Delta^{A}+\Delta^{B}\right)}. \label{eq:x1}
\end{equation}

Using the definition of the gap functions,

\begin{equation}\Delta^{\alpha}_{\bar{k}}=g^{\alpha}\left\langle c^{\alpha}_{k\sigma}
c^{\alpha}_{k-{\sigma}}\right\rangle=
\sum_{n}u_{n,\alpha}v^{*}_{n,\alpha} {\rm
th}\left(\frac{\beta{E_{n}}}{2}\right) \label{eq:gp}
\end{equation}
we can write down the gap equation for the $\Delta^{\alpha}$
\begin{align}
&\Delta^{A}=\sum_{\bar{k}}\dfrac{g^{A}}{x^{2}_{1}+x^2_2}
\left\{-x^{2}_{1}\,{\rm th}\left(\frac{\beta{E_{1}}} {2}\right)
+x^{2}_{2}\,{\rm th}\left(\frac {\beta{E_{2}}} {2}\right)\right\}, \nonumber\\
&\Delta^{B}=\sum_{\bar{k}}\dfrac{g^{B}}{x^{2}_{1}+x^2_2}
\left\{x_2^2\, {\rm th}\left(\frac {\beta{E_{1}}} {2}\right)
-x_1^2\,{\rm th}\left(\frac {\beta{E_{2}}} {2}\right) \right\}.
\label{eq:gaps}
\end{align}

Note that within our convention a positive $g$ corresponds to an
attractive on-site interaction. We now calculate $T_c$ by
linearizing the gap equations. To obtain analytical results, we
estimate the integrals involved using the approximate window
density of states (wDOS)
\begin{equation}
\rho(\omega)= \left\{\begin{array}{lr}
1/8t& \enspace  -4t\le\omega\le4t\\
\phantom{\Big|}0& \enspace \mathrm{elsewhere}\enspace ,
\end{array} \right.
\label{eq:windos}
\end{equation}
which is a good approximation for the tight binding model for
qualitative purposes. Then $T_{c}$ takes the form
\begin{eqnarray} k_{B}T_{c}&\simeq &
\left(\frac{8te^{\gamma}}{\pi}\right) \exp \left( -{\frac1{\bar
g_{eff}}}\right),
\phantom{\Big|}\nonumber\\
\bar g_{eff}&\equiv & \frac
{4g^{A}g^{B}-8t\left(g^{A}+g^{B}\right)}{8t\left(g^{A}+g^{B}\right)-
\left(8t\right)^{2}}. \label{eq:tcwdos}
\end{eqnarray}
\begin{figure}
\begin{center}
\includegraphics[width=1\columnwidth]{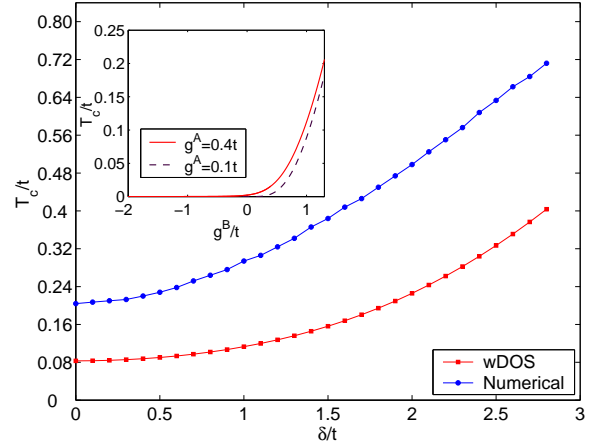}
\caption{Critical temperature $T_c/t$ plotted vs. difference of
sublattice coupling constants $\delta \equiv (g^A-g^B)/t$ for
fixed $\bar g=t$.   Squares:    window density of states. Circles:
exact. Insert: $T_c/(4t)$ vs. $g^B/t$ for fixed $g^A=t$.}
\label{fig:Tc}
\end{center}
\end{figure}

The dimensionless effective interaction $\bar g_{eff}$ plays the
same role as the single attractive coupling constant in BCS
weak-coupling theory.  The phase transition to the normal state
takes place as $\bar g_{eff}\rightarrow 0^+$, determining the
dotted line in Fig. \ref{fig:phsd}.  Note that the mathematical
expression for $T_c$ (\ref{eq:tcwdos}) actually diverges as $\bar
g_{eff}\rightarrow 0^-$, i.e. as one approaches the phase
transition from the normal state side, again in perfect analogy
with BCS. The apparent nonzero value for $T_c$ on this side is of
course spurious, as there is no ordering below this temperature,
as will be shown below.

 This estimate for $T_c$ is compared in
Figs. \ref{fig:phsd} and \ref{fig:Tc} with the exact numerical
solution for the normal state instability temperature from Eqs.
\ref{eq:gaps}. In order to judge the extent to which inhomogeneity
enhances superconductivity, in this work we keep the average
pairing interaction per site, ${\bar g}\equiv (g^A+g^B)/2$, fixed
unless otherwise stated, and plot $T_c$ vs. the difference
$\delta\equiv (g^A-g^B)$.  We see that the numerical discrepancies
between the exact evaluation and the window DOS approximation are
in fact quite large, due to the presence of the van Hove
singularity at the Fermi level in the simple 2D band, together
with the exponential dependence of $T_c$ on the density of states,
but that the approximation captures the qualitative dependence of
$T_c$ on inhomogeneity. It is clear that, as in other mean field
models of inhomogeneous pairing, the inhomogeneity enhances the
critical temperature. We return in Sec. \ref{sec:superfluid} to
the question of when this result breaks down.

If, instead of holding the average interaction $\bar g$ fixed, we
fix the attractive interaction on one sublattice $g^A$ and
progressively decrease $g^B$, we find a rapidly decreasing $T_c$
(see insert to Fig. \ref{fig:Tc}).  Depending on the magnitude of
$g^B$, we either see only an exponential decrease of $T_c$ or, for
very small $g^B$, an instability to the metallic state at a
critical value determined by the change of sign of the effective
interaction. This determines the solid line in the phase boundary
corresponding to the transition to the normal state shown in Fig.
\ref{fig:phsd}.


\section{Quasiparticle states and thermodynamics}

\subsection{Order parameters}

We now consider the  order parameters $\Delta^\alpha$ which
develop below $T_c$ on the two sublattices. It turns out that
there are   two different cases, which we discuss separately,
corresponding to coupling constants on the two sublattices  of
identical or opposite sign.
  In general, we can write using Eqs.
(\ref{eq:gaps}),

\begin{equation}
\frac{\Delta^A}{g^A}-\frac{\Delta^B}{g^B}=\sum_{\bar{k}}
\left[ \tanh\left(\frac{\beta{E_2}}{2} \right) -
\tanh\left(\frac{\beta{E_1}}{2}\right) \right]
\label{eq:gapdiff}
\end{equation}

We first assume both coupling constants positive, $0\le g_B\le
g_A$.  If both gaps are positive, then $E_{1},E_{2}$ are positive.
Therefore in limit $T\rightarrow0$ the right hand side of Eq.
(\ref{eq:gapdiff}) will vanish, which gives the solution:

\begin{equation}
\frac{\Delta^{A}}{\Delta^{B}}=\frac{g^{A}}{g^{B}},
\label{eq:gapratio}
\end{equation}
which  is exact for the case of both $g's$ attractive. Now if we
employ the window density of states given by Eq.
(\ref{eq:windos}), we obtain

\begin{equation}
\Delta^{\alpha}\simeq\frac{g^{\alpha}}{g^{A}+g^{B}}\frac{8t}{\sinh\left(\frac{8t}{g^{A}+g^{B}}\right)}
\label{eq:gapwindos}
\end{equation}
For $g^{A}=g^{B}$, within weak coupling approximation
($g^{A,B}<<8t$), this gives $\Delta/k_{B}T_{c}=e^{\gamma}/\pi$,
which is the standard BCS result.

\begin{figure}
\begin{center}
\includegraphics[width=1.02\columnwidth]{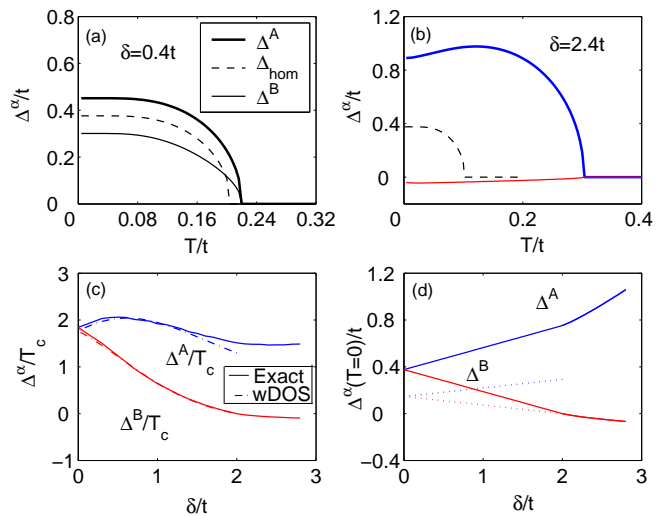}
\caption{(a) Sublattice order parameters $\Delta^{A,B}$ vs. $T/t$
for weak inhomogeneous pairing interaction $\delta\equiv
(g^A-g^B)=0.4t$; (b)  strong inhomogeneity cases, with
$\delta=2.4t$ (note $g^B=-0.2t$).  (c) Ratio
$\Delta^\alpha(T=0)/T_c$  vs. $\delta$ for $\alpha=A,B$ (solid
line: full numerical evaluation; dashed-dotted line: wDOS); (d)
Comparison of full numerical evaluation of $\Delta^\alpha(T=0)$
(solid line) with wDOS approximation (dashed-dotted line). For all
plots, the average coupling is fixed to $\bar g=1$.  Note in
(c),(d) that the gapless phase discussed in Sections
\ref{sec:condensate} and \ref{sec:dispersion} correspond, for the
parameters given here, to $2\le \delta/t$. }
\label{fig:gaps_vs_T}
\end{center}
\end{figure}
\noindent The exact solution for the temperature dependence of the
gap is shown for the case of weak inhomogeneity in Fig.
\ref{fig:gaps_vs_T}(a).

Now  we consider the other case when one of the coupling constants
is negative, $g_B\le 0 < g_A$.  From (\ref{eq:gaps}) we see that
 $\Delta^B$ is forced to be negative.  The variation of gap as a
function of temperature found numerically and displayed in Fig.
\ref{fig:gaps_vs_T}(b) is somewhat unusual.  As the temperature
increases from zero, the gap increases in magnitude and then after
attaining a maximum value begin to decrease.


In the case of opposite sign coupling constants, relation
(\ref{eq:gapratio}) is no longer valid. 
In this case one can obtain an approximate analytical expression
for $|\Delta^B|/\Delta^A$ using the wDOS (\ref{eq:windos}), but it
is quantitatively inaccurate.  A more accurate approximation is
obtained by replacing the exact contour of zero quasiparticle
energy with a square of side
$2\sqrt{2}\cos^{-1}\left(\frac{\sqrt{\Delta^{A}
\left|\Delta^{B}\right|}}{4t}\right)$.  This turns out to be
equivalent to a renormalized wDOS approximation where the bare
bandwidth $8t$ is replaced by $4\pi t$.  We find
\begin{equation}
\frac{\left|\Delta^B\right|}{\Delta^A}=\left(\dfrac{g^B}{2\pi t}\right)^2
\left(\sqrt{\dfrac{4\pi^2t^2}{g^A\left|g^B\right|}+1}-1\right)^2 \, .
\label{eq:ngrat}
\end{equation}



In Figs. \ref{fig:gaps_vs_T} (c) and (d) we illustrate the utility
of the wDOS and renormalized wDOS approximations with which
analytical results were obtained above.  While Fig.
\ref{fig:gaps_vs_T}(d) shows that the inaccuracies arising from
the wDOS also influence the value of the $T=0$ order parameters on
the two sublattices, the qualitative tendencies with $\delta$ are
well reproduced. Furthermore, the ratios $\Delta^{A,B}/T_c$ agree
quantitatively with the exact numerical results over a wide range
of inhomogeneities (Fig. \ref{fig:gaps_vs_T}(c)), as the errors
introduced in the DOS are largely cancelled by taking the ratio.




The normal state region shown in the phase diagram of Fig.
\ref{fig:phsd} represents the  metallic phase where, within the
current mean field treatment, the only solution found has both
order parameters $\Delta^{A,B}=0$.

\subsection{Condensate amplitudes}
\label{sec:condensate} 
\begin{figure}[t]
\includegraphics[width=1.1\columnwidth]{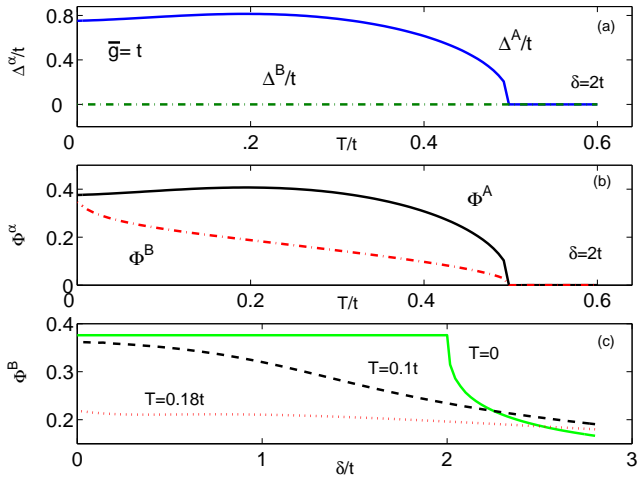}
\caption{(a) Gaps $\Delta^{A,B}/(4t)$ vs. $T/(4t)$ for special
case $\delta=2t$, $\bar g$=t $(g^B=0)$; (b) Condensate amplitude
$\Phi^B\equiv \langle c^{B}c^{B}\rangle $ for same case as (a);
(c) Variation of $\Phi^B$ with inhomogeneity $\delta/t$ for three
different temperatures $T/t=0,0.11,0.18t$.} \label{fig:condensate}
\end{figure}

There are several measures of the
``strength of superconductivity" in a material which become
distinct when the material is inhomogeneous.  One is the order
parameter, which we have discussed above.  A second is the pair
wave function, or condensate amplitude $\Phi^\alpha \equiv \langle
c^\alpha_{-\k-\sigma} c^\alpha_{\k\sigma}\rangle$, such that
$\Delta^\alpha = g^\alpha \Phi^\alpha$.   To illustrate the
differences, it is useful to consider first an interesting if
artificial special case where the coupling is attractive on one
sublattice and vanishes on another. We therefore take $g^A=2t$ and
$g^B=0$, and plot both the order parameters and condensate
amplitudes in Fig. \ref{fig:condensate}.

%

Using the  wDOS, the condensate on either the A or B sublattice
for the $g^B=0$ case can be obtained analytically at $T=0$ as
\begin{equation}
\Phi^\alpha= 4 \sinh^{-1}\left[\frac{1}{2\sinh(8t/g^A)}\right]
~~~(\alpha=A,B, ~~g_B=0)\label{eq:condensate}
\end{equation}
In other words, although the pairing interaction on the B
sublattice is zero, the proximity effect is so strong that a
uniform condensate fraction is induced over the entire system.
This degeneracy is broken at finite temperatures, where the B
condensate is weakened relative to that on the A sublattice.  Of
course, the order parameter on the B sublattice is zero at all
temperatures due to the absence of coupling on these sites.  These
differences are illustrated in Fig. \ref{fig:condensate}(a) and
(b). It is also interesting to study the dependence of the
subdominant condensate on inhomogeneity.  The point where one of
the coupling constants changes sign is clearly a singular point of
the theory, such that at $T=0$ this condensate is immediately
depleted as soon as the subdominant coupling (B sublattice)
becomes negative.  We now discuss the reasons for this abrupt
change in behavior.

\subsection{Dispersion and density of states}
\label{sec:dispersion}
\begin{figure}
  \begin{center}
  \includegraphics[width=1.05\columnwidth]{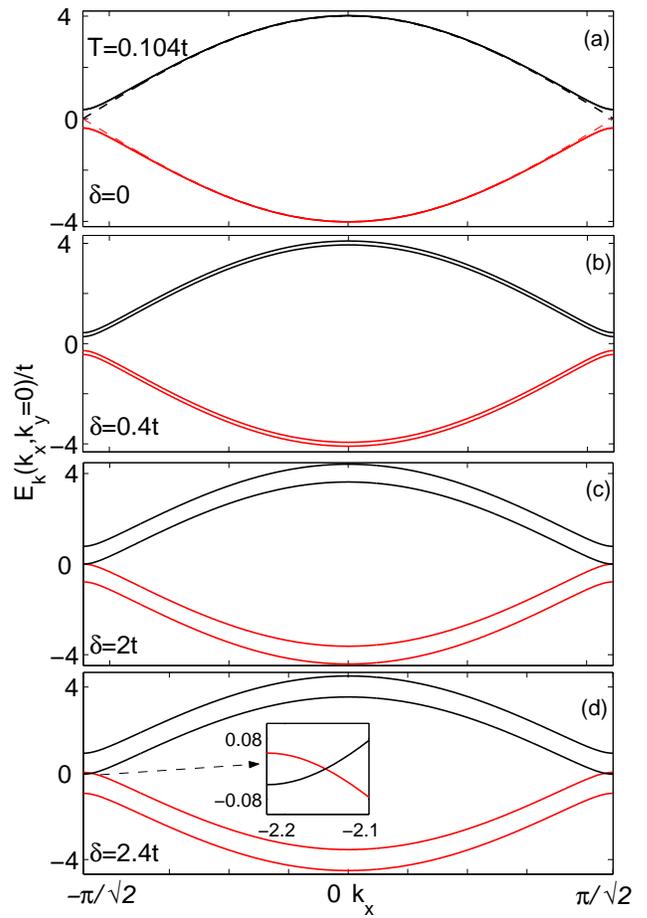}
  \caption{The quasiparticle energy (in units of $t$) in momentum
space along $k_{x}$ (on x- axis),  with $k_{y}=0$ for average
coupling $\bar g=t$ and various values of inhomogeneity
$\delta=g^a-g^b$. (a)Spectrum when $g^A=g^B$.  Dashed curve
represents normal state $g^A=g^B=0$ for reference. (b) Spectrum
for $\delta=0.4t$ (c) Spectrum for $\delta=2t$, at transition to
gapless state. (d) Spectrum for $\delta=2.4t$, in gapless state.
All spectra are shown at temperature $T=0.104t$. }
\label{fig:spectra}
  \end{center}
\end{figure}

In Eqs. (\ref{eq:E1}), we have given the quasiparticle
eigenenergies for this simple model;   we now plot them in Fig.
\ref{fig:spectra} along $k_x$ for various representative cases. Of
course the full quasiparticle spectrum retains the fourfold
symmetry of the underlying lattice. Panel a) shows the dispersion
in both the normal and homogeneous BCS case, respectively.  Adding
a small inhomogeneity with two attractive couplings leads to a
splitting of these states, into
 four distinct eigenenergies (panel (b)), thereby
reducing the overall spectral gap in the system.  When one of the
couplings passes through zero, a level crossing occurs at zero
energy at the zone face (panel (c)).  It is this level crossing
which is responsible for the singular behavior in some observable
quantities as one of the couplings becomes repulsive.  It is
accompanied by a filling of the gap by the states near the zone
face.  When the two couplings $g_A$ and $g_B$ have opposite signs,
the eigenvalues change sign over the Brillouin zone. For example,
in panel (d) of Fig. \ref{fig:spectra}, $E_{2}$ is always
positive, but $E_{1}$ is positive except in a narrow range
$-\sqrt{\Delta^{A}\left|\Delta^{B}\right|}\le\xi\le\sqrt{\Delta^{A}\left|\Delta^{B}\right|}$.

Gapless behavior in $s$-wave superconductors with magnetic
impurities\cite{Abrikosov61}
 can be associated with individual low-energy quasiparticle resonances,
 which arise near each magnetic impurity and
 then overlap to form an impurity band.
 The band appears as a residual density of states
 for sufficiently high impurity concentration
Similarly, the gapless superconductivity arising due to
microscopically inhomogeneous sign-changing order parameters,
which is found in the present paper, is reminiscent of respective
low-energy Andreev  states around so-called off-diagonal or phase
impurities \cite{Chattopadhyay,Andersen_phaseimpurity}. One might
imagine an ordinary $s$-wave superconductor into which a dilute
lattice of phase impurities is introduced.  As the lattice
constant of such bound states is decreased, the bound state wave
functions will interfere and the bound state energies will split,
leading to a kind of gapless superconductivity related to that we
find here for the dense case.

\begin{figure}
\begin{center}
\includegraphics[width=1\columnwidth]{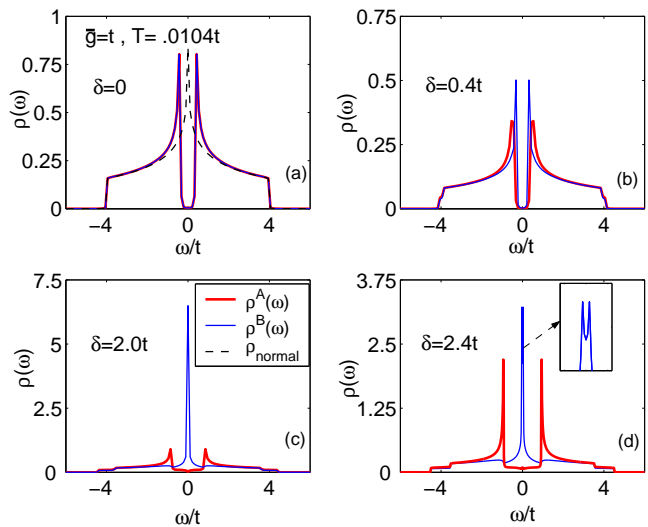}
\caption{Local density of states (LDOS) at sites $A$ (red) and $B$
(blue). (a) LDOSin homogeneous case, $\delta=0$;  (b)
$\delta=0.4t$, $\bar{g}=t$;  (c) $\delta= 2t$, $\bar{g}=t$;  (d)
$\delta=2.4t$, $\bar{g}=t$. The dashed line represents the normal
state 2D tight binding band. }\label{fig:dos}
\end{center}
\end{figure}

To observe this directly and trace the spectral features back to
their respective sublattices, we construct the local density of
states on A and B sites, \begin{equation}\rho^\alpha(\omega) =
\sum_{n,\k} \, \left[|u_{n}^\alpha|^2 \delta(\omega-E_n) +
                                          |v_{i}^\alpha|^2\delta(\omega+E_n)
                                          \right].
\end{equation} 
%
%

Fig. \ref{fig:dos} exhibits the LDOS on the two sublattices.  In
panels (a) and (b), when both couplings are positive, we clearly
see a full spectral gap. But when one of the couplings tends to
zero, as in (c), a sharp peak develops at the Fermi level on the
associated sublattice.  The system remains gapless as the coupling
constant on this sublattice is made negative, as in (d).  It
should be noted, however, that the subgap states are not located
at zero energy, as seen in Fig. \ref{fig:spectra}d, where the
dispersion is seen to flatten at the zone edge away from zero.
This is reflected in a narrowly split resonance in the LDOS, as
seen in Fig. \ref{fig:dos}.


\begin{figure}[t]
\begin{center}
\includegraphics[width=0.9\columnwidth]{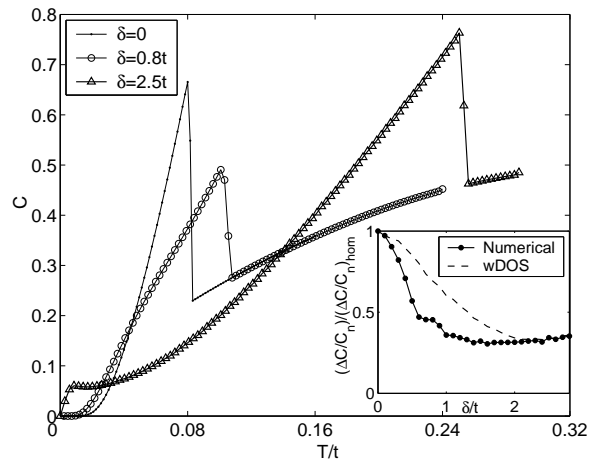}
\caption{Specific heat $C(T)/T$ of inhomogeneous system vs.
$T/(4t)$.
 Solid line: BCS homogeneous superconductor with 2D tight-binding band.
 Circles: weak homogeneity $\delta=0.8t$.  Triangles: strong
 homogeneity $\delta=2.5t$.  Average coupling fixed to $\bar g=t$.
 } \label{fig:dcv}
\end{center}
\end{figure}
\subsection{Specific heat} To exhibit the consequences of the
sublattice pairing inhomogeneity for thermodynamics, we calculate
the specific heat from the temperature derivative of the entropy,
\begin{equation}
C=T\frac{dS}{dT}.
\end{equation}
Here $S$ for the free quasiparticles is given by
\begin{equation}
S= \sum_{\bar{k},n=1,2}\left(f_{n}
\log\left(f_{n}\right)+\left(1-f_{n}\right)
\log\left(1-f_{n})\right)\right), \label{eq:scv}
\end{equation}
where $f_n\equiv f(E_n(\bf k))$, and the sum runs only over
positive eigenvalues. In Figure \ref{fig:dcv}, we plot these
results for a few examples. For a superconductor with a small
amount of inhomogeneity--again keeping the average coupling
fixed--it is clear that the critical temperature increases
relative to the BCS case,  the low-$T$
 temperature dependence remains exponential.  On the other hand, when one of the coupling
 constants goes negative, the figure shows the existence of the
 residual density of states via a linear term in the specific heat
 at the lowest temperatures.

It is clear  that the
 fractional jump in the specific heat, reflecting the condensation
 energy, decreases with increasing inhomogeneity, {{and is approximately constant for the gapless regime.}}
 Note that this ratio
for the homogeneous case is not the conventional BCS value of
$1.43$,  due to the non-parabolic band used here.

\section{Superfluid density}\label{sec:superfluid}

  As Aryanpour et al. have
emphasized \cite{scalettar}, models of inhomogeneous pairing can
lead to a variety of ground states, including insulating ones. To
show that the   states we discuss here are indeed superconducting,
we
 use the criteria developed by
Zhang et. al. \cite{scal} for lattice systems.  We calculate the
superfluid density $n_s$, which is the sum of  the diamagnetic
response of the system  (kinetic energy density) and paramagnetic
response  (current-current
 correlation function)
\begin{equation}
n_s/m = \langle -k_x \rangle - \Lambda(\omega=0,\q\rightarrow 0).
\label{eq:rhos_twoterms}
\end{equation}
The current-current response  is defined as
\begin{equation}
\Lambda_{xx}(\bar{q},i\omega_m)=\int^\beta_0 d\tau
 e^{i\omega_m \tau}\left\langle
 j^{P}_{x}(\bar{q},\tau)j^{P}_{x}(-\bar{q},0)\right\rangle
\label{eq:currentcorr}
\end{equation}
 where the current at $i^{th}$ site is given by
 \begin{equation}
 j^{P}_{x}(i)=i t
 \sum_{\sigma}\left(c^{\dagger}_{i+x,\sigma}c_{i,\sigma}
 -c^{\dagger}_{i,\sigma}c_{i+x,\sigma}\right).
 \label{eq:current}
\end{equation}

After evaluating the expectation value using the Bogoliubov
operators in (\ref{eq:ctran}), we find the following
relatively simple expression for the analytic continuation of the
static homogeneous response:

\begin{eqnarray}
\Lambda_{xx}(\omega=0,\q\rightarrow 0)&=&2 \sum_{k} \left[4t
\sin(\frac{k_x}{\sqrt{2}})\cos(\frac{k_y}{\sqrt{2}})\right]^{2}
 \\ &&\times \left(\frac{f(E_1)-f(E_2)}{E_1-E_2}\right) \nonumber  \label{eq:cucuco}
\end{eqnarray}

If the $E_i(\k)$ do not change sign over the Brillouin zone, as is
the case for the gapped phase $g_A,g_B>0$, it is clear that this
expression vanishes as $T\rightarrow 0$ as in the clean BCS case.
This is no longer the case in the gapless regime $g_B\le 0$, where
a finite value of $\Lambda$, corresponding to a  residual density
of quasiparticles,  is found at zero temperature.

\begin{figure}[t]
\includegraphics[width=0.48\textwidth]{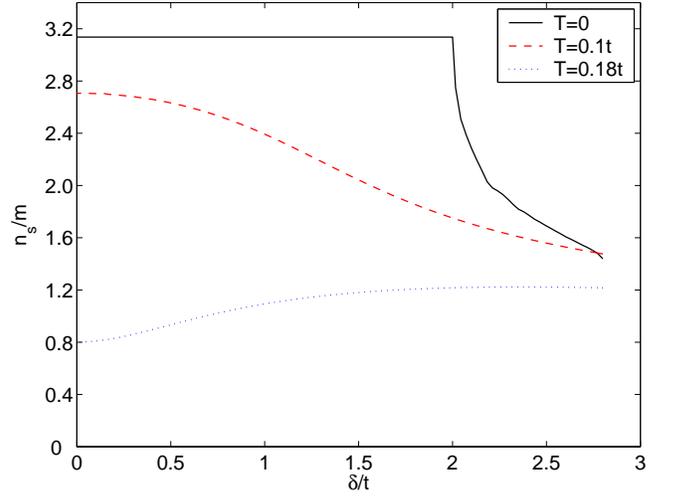}
\caption{Superfluid density $n_s$ vs. inhomogeneity $\delta$ with
fixed average coupling $\bar g = t$ at different temperatures
$T=0$ (solid) $0.1t$ (dashed), and  $0.18t$ (dotted line).  The
range of inhomogeneities  $2\le \delta/t $ corresponds
to gapless superconductivity. } \label{fig:kx}
\end{figure}

 The other term we need to evaluate is the expectation value of the lattice kinetic energy density operator,
 which
  in the homogeneous  BCS case is directly proportional to the superfluid
 weight $n_{s}/m=\langle-k_{x}\rangle$ at $T=0$. The kinetic energy operator is
 defined as,
\begin{equation}
k_{x}(i)=-t\sum_{\sigma}\left(c^{\dagger}_{i+x,\sigma}c_{i,\sigma}
+c^{\dagger}_{i,\sigma}c_{i+x,\sigma}\right)
\label{eq:kx}
\end{equation}
In Fourier space, it can be written as:
\begin{equation}
k_{x}(i)=-t\sum_{\bar{k_{1}},\bar{k_{2}},\sigma}\left(c^{A\dagger}_{\bar{k_{1}}}
 c^{B}_{\bar{k_{2}}}e^{i\left(\bar{k_{2}}-\bar{k_{1}}\right)\cdot
 \bar{r_{i}}}\left[2 \cos\left(\frac{k_{2y}}{\sqrt{2}}\right)
 e^{i\frac{k_{2x}}{\sqrt{2}}}\right]+h.c.\right)
\label{eq:kxf}
\end{equation}
After performing the canonical transformation and taking the
 expectation values we find \vskip 0cm
\begin{multline}
\left\langle {k_{x}(i)}\right\rangle =
-2t \sum_\k\cos\left(\frac{k_{y}}{\sqrt{2}}\right)\times\\
\times\sum_{n}\left\{\left[u^{A*}_{n}u^{B}_{n}
 e^{i\frac{k_{x}}{\sqrt{2}}}+u^{A}_{n}u^{B*}_{n}
 e^{-i\frac{k_{x}}{\sqrt{2}}}\right] f(E_{n})+ \right.\\ \left.
+\left[v^{A}_{n}v^{B*}_{n}e^{i\frac{k_{x}}{\sqrt{2}}}+
 v^{A*}_{n}v^{B}_{n}e^{-i\frac{k_{x}}{\sqrt{2}}}\right]
 \left(1-f(E_{n})\right)\right\}
\label{eq:expkx}
\end{multline}
which is, remarkably, the same for  sites A or B, corresponding to
the uniform proximity-induced condensate amplitude found in
section \ref{sec:condensate}. In terms of the eigenvectors and
 eigenvalues given by Eqs. (\ref{eq:v1}), (\ref{eq:v2}),
 (\ref{eq:v3}) and (\ref{eq:v4}), we can write
\begin{multline}
\left\langle{k_{x}(i)}\right\rangle =2\sum_{\bar{k}} \xi_\k
\frac{x_{1}x_{2}}{x_{1}^{2}+x_{2}^{2}}
\left[\tanh\left(\frac{\beta{E_{1}}}{2}\right)+\right.\\ \left.
\tanh\left(\frac{\beta{E_{2}}}{2}\right)\right] \label{eq:fnkx}
\end{multline}

For the case when $g^{A},g^{B}>0$ at $T=0$, we can simplify this
 expression as
\begin{equation}
\left\langle{-k_{x}(i)}\right\rangle=2\sum_{\bar{k}}
 2\frac{\xi_\k^{2}}{\sqrt{\left(\Delta^{A}_{0}+\Delta^{B}_{0}\right)^{2}+4\xi_\k^{2}}},
\label{eq:kxt0}
\end{equation}
which  is manifestly positive; hence,
  the system is a superconductor and displays a conventional Meissner effect at $T=0$.
The expression also shows  explicitly that the superfluid density
on each site corresponds to the average superconducting order
parameter over the system.

\begin{figure}
\includegraphics[width=0.5\textwidth]{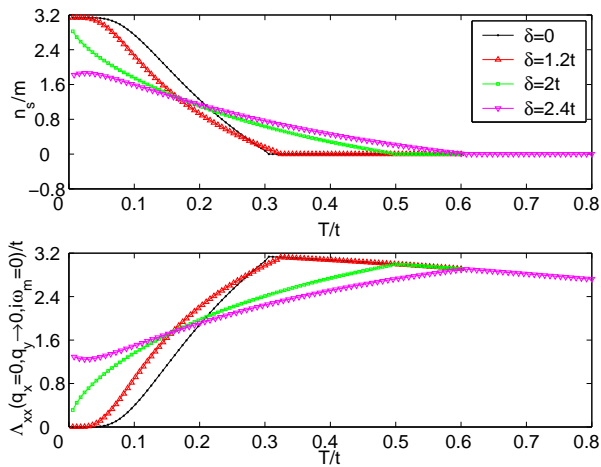}
\caption{The upper graph shows the superfluid density $n_s$ as a
function of temperature for various inhomogeneity parameters
$\delta/t$= 0 (black), 1.2 (red), 2 (green) and $\bar g= t$
(magenta). The lower panel shows the current current correlation
$\Lambda_{xx}(T)$.} \label{fig:rhovsT}
\end{figure}

Fig. \ref{fig:kx} show how the superfluid weight $(n_s)$ changes
as the inhomogeneity is increased.  At $T=0$, $n_s$ for this model
is a constant and equal to the value for the homogeneous system,
whenever the system is fully gapped, since the average gap remains
the same.  In the same system, recall, the transition temperature
increases monotonically with inhomogeneity. As one increases the
inhomogeneity further with fixed average coupling, there is a
critical value $\delta_c$ for which the system enters the gapless
phase, at which the superfluid density drops abruptly. The
temperature dependence of such a case is exhibited, along with
others, in Fig. \ref{fig:rhovsT}, and we see that this
discontinuity corresponds to the creation of a finite residual
normal fluid of uncondensed quasiparticles.  Exactly at $\delta_c$
($=2t$ in Fig. \ref{fig:rhovsT}), this behavior is  marginal, but
for any $\delta>2t$ in the Figure, the residual normal fluid
density $\Lambda_{xx}(T=0)$ is nonzero.

\begin{figure}[t]
\includegraphics[width=0.48\textwidth]{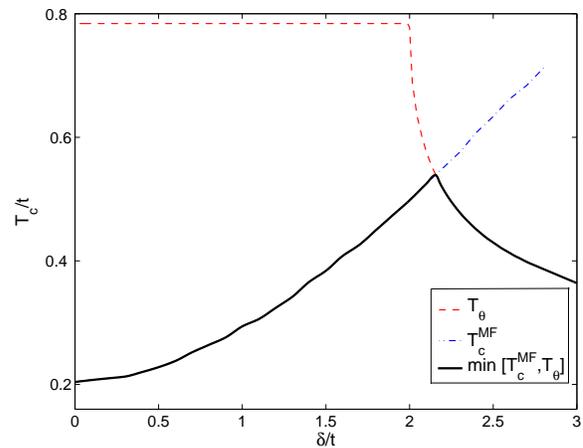}
\caption{Mean field critical temperature $T_c$ (dashed-dotted
line) and phase ordering temperature $T_\theta$ (dashed line) vs.
inhomogeneity $\delta$ for fixed average interaction $\bar g=t$.
The solid line is the minimum of the two curves at any $\delta$.}
\label{fig:optimalTc}
\end{figure}

If $\bar g$ is held fixed and inhomogeneity $\delta$ increased
further, there is no  mean-field transition to the metallic state,
as indicated e.g. by the dashed-dotted line in Fig. \ref{fig:phsd}
which we have investigated in this paper.  Instead, $T_c$
increases monotonically while the superfluid density begins to
fall at $T=0$ when the system enters the gapless phase at
$\delta_c$. Eventually, the decrease in $n_s$ must lead to large
phase fluctuations and the breakdown of the mean field theory
described here.  To estimate when this occurs, we follow Emery and
Kivelson \cite{EmeryKivelson} and define a characteristic phase
ordering temperature
\begin{equation}
T_\theta = \hbar^2{n_s(0)\ell/4m^*},
\end{equation}
which will now decrease in the gapless phase, following the
superfluid density. For quasi two-dimensional superconductors, the
length scale $\ell$  is a larger of the average spacing between
layers $d$ or $\sqrt{\pi}\xi_{\perp}$, where $\xi_{\perp}$ is the
superconducting coherence length perpendicular to the layers.
Although a determination of $\ell$ is beyond our simple model, it
is qualitatively irrelevant in the presence of significant abrupt
drop of the superfluid density. Taking simply $\ell\sim a$, which
is within a factor of 2-3 of both $d$ and $\sqrt{\pi} \xi_\perp$
in the cuprates, we may now obtain a crude measure of the {\it
optimal inhomogeneity} needed to maximize the true ordering
temperature in this model. In Fig. \ref{fig:optimalTc}, we now
plot both the mean field $T_c$ and the putative phase ordering
temperature as a function of inhomogeneity $\delta$.  We see that
the two curves cross very close to the phase boundary $\delta_c$
(at $\delta=2t$ in the Figure), due to the steep drop in $n_s$
there. It therefore appears that the optimal inhomogeneity {\it
within the current scheme} occurs for a checkerboard-like pairing
interaction with attractive interaction on one sublattice and zero
interaction on the other.

\section{Conclusions}

We have presented and solved a simple mean-field model of a
superconductor with inhomogeneous  pairing varying on an atomic
scale.  The model simply takes a different value for pairing on
each of two interpenetrating sublattices.  The inhomogeneity can
be increased while keeping the average pairing in the system
fixed, thus allowing the effect of inhomogeneity on various
properties to be discussed.  Many results for this model were
presented in analytic form.   As pointed out by earlier workers,
we find that increasing pairing inhomogeneity enhances $T_c$.  We
have presented analytical results reflecting this effect, and
calculated the spectrum and thermodynamic properties as well. The
enhancement of $T_c$ is accompanied by a closing of the spectral
gap and eventually by the appearance of gapless phase in the
quasiparticle spectrum of the ``$s$-wave" superconductor. While
these results have been obtained at the mean field level, because
the length scale of the variation of the pairing interaction is so
short we do not expect fluctuations to change them significantly
\cite{Kivelson}, at least until the superfluid density is
substantially suppressed.

The last caveat does not seem to be particularly restrictive,
however.   The enhanced $T_c$ found in the inhomogeneously paired
system is accompanied by a reduction in condensation energy, but
the system maintains a constant $T=0$ superfluid density until a
critical value of the inhomogeneity, when a level crossing of two
quasiparticle energies forces the system into the gapless phase.
Until this point, the condensate at $T=0$ is spatially homogeneous
for any value of the interaction inhomogeneity, even when the
interaction on a given sublattice is zero; this reflects the
strong proximity coupling between the sublattices.  As
inhomogeneity increases further, we have argued that the decrease
in the superfluid density leads fairly rapidly to a decrease in
the phase ordering temperature; we find, therefore, that the
optimal inhomogeneity is reached close to the checkerboard pattern
of interactions, with attractive pairing on one sublattice and
zero on the other.

Our results  suggest that in superconducting systems with short
coherence length, a modulated pairing interaction at the atomic
scale may provide a route to high temperature superconductivity.

The authors are grateful for useful conversations with S. A.
Kivelson.  Research was partially supported by DOE
DE-FG02-05ER46236 (PJH and VM) and RFBR grant 08-02-00842 (YSB).

\appendix

{}

\end{document}